# Magnetocaloric properties of $La_{0.7}Ca_{0.3}Mn^{16}O_3$ and $La_{0.7}Ca_{0.3}Mn^{18}O_3$ manganites and their 'sandwich'

A.M. Aliev[1, a)], A.G. Gamzatov[1], K.I. Kamilov[1,2], A.R. Kaul[3], N.A. Babushkina[4]

[1] Amirkhanov Institute of Physycs of Daghestan Scientific Center, RAS, Makhachkala 367003, Russia

[2] Department of Physics, Moscow State University, Moscow 119899, Russia

[3] Department of Chemistry, Moscow State University, Moscow 119899, Russia

[4] Institute of Molecular Physics, Russian Research Center "Kurchatov Institute", Moscow 123182, Russia

The effect of $^{16}O \rightarrow ^{18}O$ isotope substitution on specific heat and magnetocaloric effect of polycrystalline $La_{0.7}Ca_{0.3}MnO_3$ manganite is studied. Mainly the effect of isotope substitution for the specific heat and magnetocaloric effect is only the reduction of temperatures of anomalies. $\Delta T_{ad}$ values at magnetic field change $\Delta H$ = 18 kOe are equal $\Delta T_{ad}$=2.41 K and 2.60 K for LCMO16 and LCMO18 respectively. The sandwich of the LCMO16 and LCMO18 samples was produced for direct measurement of $\Delta T_{ad}$ The use of sandwich from materials with near similar magnetocaloric properties increases the RCP by about 20%.

[a)] Electronic mail: lowtemp@mail.ru

Magnetocaloric effects at second-order phase transitions are usually observed in a wide range of temperatures above and below the transition point, but its magnitude, except for Gd, is not sufficient for applications in magnetic cooling. Materials with coupled magnetostructural first-order phase transitions are considered to be more promising in magnetic cooling technology [1-3]. Due to the abrupt change of lattice parameters at these transitions, a sharp change of magnetization has been observed. The Maxwell relation implies that for such transitions, the magnitude of the MCE can reach very high values. In addition, the contribution to overall MCE can be given by the entropy change associated with the structural transition. However, at the same time temperature region where the MCE magnitude reaches significant values is quite narrow. Moreover, the effectiveness of these materials as a working agent for magnetic refrigerators might also be decreased due to the temperature hysteresis typical for the first-order phase transitions. Having multiple materials with large MCE, but a small temperature width of the effect, one can make 'sandwiches' from these materials to increase the relative cooling power (RCP). However, the proximity of the critical temperatures of raw materials at preparation of 'sandwiches' is prerequisite for increasing the RCP. One of the materials with large MCE is the $La_{1-x}Ca_xMnO_3$ manganite. Magnetic phase transitions in these manganites are observed near room temperature for concentrations $x \geq 0.3$, and the values of the MCE reach significant magnitudes [4-8]. These materials are studied in various crystalline forms, including the nanocrystalline ones [8-10].

The order of the phase transition in $La_{0.7}Ca_{0.3}MnO_3$ is still under consideration. Some authors attribute this transition to the first-order ones [5, 11], others considered to be second-order transition [12]. Clear evidence for the first order nature of the phase transition in LCMN is provided in [13]. It is well known that indirect methods for MCE evaluation (by measuring the magnetization and specific heat) using the Maxwell or thermodynamic relations ratio for the first-order phase transitions often give the wrong, mostly overestimation values. So, for the determination of the magnetocaloric properties of these materials a direct measurement of MCE is required.

In this work we studied the effect of $^{16}O \rightarrow ^{18}O$ isotope substitution on specific heat and magnetocaloric effect of polycrystalline samples of $La_{0.7}Ca_{0.3}Mn^{16}O_3$ (LCMO16) and $La_{0.7}Ca_{0.3}Mn^{18}O_3$ (LCMO18) manganites. The method of $^{16}O \rightarrow ^{18}O$ isotope substitution is described elsewhere [14]. The oxygen stoichiometry preserved during the isotope substitution and strictly controlled. The oxygen stoichiometry of LCMO16 sample was determined by iodometric titration. The $^{18}O$ content in the sample was determined by measurement of the weight change of the sample after the isotope substitution. As a result, the completeness of the isotope exchange was verified. Samples under studying were plates with approximate sizes of 3x3x0.3 $mm^3$. Direct measurements of $\Delta T_{ad}$ were carried out both by classical method, when temperature change under external magnetic field is measured, and by the modulation method [15]. In the last case the sample is subjected to alternating magnetic field, which induces a periodic change in temperature of the sample due to magnetocaloric effect. This temperature variation through a differential thermocouple, one junction of which is glued to the test sample, is registered with lock-in amplifier. Magnetic field modulation frequency in experiment was 0.3 Hz. Alternating magnetic field up to 4 kOe was generated using an electromagnet and a power supply with an external control. Control voltage was produced by generator of the lock-in amplifier. Alternating magnetic field 18 kOe was created using Adjustable Permanent Magnet Based Magnetic Field System (AMT & C LLC, Troitsk, Russia). To compare with direct measurements, isothermal magnetic entropy change $\Delta S_m$ was calculated from specific heat in and without magnetic field. Specific heat and MCE were measured on the same samples and the same mounting of the samples.

Figure 1 shows the temperature dependence of specific heat LCMO16 and LCMO18 in the magnetic field of 18 kOe and without field, which are well manifested anomalies caused by the paramagnetic - ferromagnetic transition. Peaks of anomalies at zero-field are observed at temperatures $T_C$ = 257.3 and 249.1 K for LCMO16 and LCMO18 respectively. External magnetic field $H$ = 18 kOe significantly suppresses anomalies and shifts temperature of their peaks to 270.3 and 259.7 K, respectively. Specific heat anomalies at zero-field reach significant values, and the temperature width of the transitions is quite narrow. Such specific heat behavior is typical for the first-order phase transitions, which is confirmed by most results in other papers [4-5, 11]. Transition entropy for LCMO16 and LCMO18 are 2.31 and 2.36 J/mol K respectively, i.e. in fact equal to each other within the experimental error. As can be seen, the effect of $^{16}O \rightarrow {}^{18}O$ isotope substitution for the specific heat of $La_{0.7}Ca_{0.3}MnO_3$ is only the reduction of the critical temperature $T_C$ for ~ 8 K.

Figure 2 shows the results of direct measurements of the adiabatic temperature change $\Delta T_{ad}$ in weak magnetic fields. MCE values at $\Delta H$=1000 Oe for both samples is approximately the same, and is equal $\Delta T_{ad} \approx 0.19$ K, at T=257.6 K and 247.5 K for LCMO16 and LCMO18 respectively. At $\Delta H$ = 3500 Oe the MCE value is much larger and is equal $\Delta T_{ad} \approx 0.78$ K and 0.87 K for LCMO16 and LCMO18 respectively. These results indicate that the effect of isotopic $^{16}O \rightarrow {}^{18}O$ substitution on the magnetocaloric properties of $La_{0.7}Ca_{0.3}MnO_3$, as for specific heat, is only in the shifting of the temperature range of effect to lower temperatures. No significant difference in the magnitude or the behavior $\Delta T_{ad}(T)$ is observed. It should be noted that the value of MCE for $La_{0.7}Ca_{0.3}Mn^{18}O_3$ at low fields is quite significant, and at $\Delta H$ = 3500 Oe $\Delta T_{ad}$ is only 1.5 times less than in gadolinium at the same change of magnetic field.

Adiabatic temperature change $\Delta T_{ad}$ values at magnetic field change $\Delta H$ = 18 kOe are much larger (Fig. 3), the maximum values are $\Delta T_{ad}$=2.41 K and 2.60 K for LCMO16 and LCMO18 respectively. Temperature maxima at the same time are significantly shifted towards higher, $T_{max}$=263.7 and 254.3 K for LCMO16 and LCMO18 respectively. The values of MCE obtained in this paper are among the largest for manganites. There are many papers with results that are close or even exceed these values, but almost all of them were obtained using indirect methods, or by extrapolation of the low-field data [4]. The important feature for use in magnetic refrigeration technology is the symmetry of MCE curve above and below the temperature of the maximum. For the studied samples, such symmetry is almost perfect both at low and at moderate magnetic fields. However, keeping in mind the advantages of these materials, it must be noted that the full width of half-maximum of the effect is very small and is about 15 K for both samples. Accordingly, the relative cooling power RCP will be small also.

Figure 4 shows the results of magnetic entropy change calculation from specific heat data using the thermodynamic relation $\Delta S_m = \int_{T_1}^{T_2} \left( C_P(T, H_0) - C_P(T, H_1)/T \right)_{P,H} dT$. The behavior of magnetic entropy change $\Delta S_m$ is similar to that of $\Delta T_{ad}$. The maximum values of magnetic entropy changes are $\Delta S_m$= 5.91 and 6.05 J/kg K at temperatures $T_{max}$ = 261 and 253.2 K for LCMO16 and LCMO18 respectively. For comparison, the $\Delta S_m(T)$ curves calculated from the specific heat $C_p(H)$ and $\Delta T_{ad}$ data using a relation $\Delta S_m = \frac{\Delta T_{ad}}{T} C_P(H)$ at 18 kOe field is also shown at fig 4. As can be seen, $\Delta S_m$ values obtained by different methods agree quite well, except for the temperature shift of the peaks. $\Delta S_m$ peaks are observed at the crossing point of the specific heat in the field and without the field curves. Figures 3 and 4 indicate that the $\Delta T_{ad}$ peaks are shifted toward higher temperatures as compared with the $\Delta S_m$ peaks (according to the $C_p$ in and without field). Usually the MCE peak temperature depends weakly on the magnetic field. Here we see the shift of the peaks relative to $T_C$ (as $T_C$ we can approximately take the temperature of specific heat maximum in zero magnetic field) is about 6 K for both compounds in 18 kOe field. This temperature shift can be a consequence of using the sinusoidal magnetic field $H = H_{max}\sin(\omega t)$ for MCE measuring by modulation method. I.e. we can assume that the sample is in constant effective field produced by an alternating magnetic field. Here this field is $H_{eff} = \frac{H_{max}}{\sqrt{2}} = 13$ kOe. This field maintains a magnetically ordered state up to higher temperatures, and thus shifts the temperature of anomalies, including the MCE. Since in real magnetic refrigerators working medium will rotate in a magnetic field, i.e. actually be exposed to an alternating magnetic field, in the construction of refrigerators one needs to take into account this shift of the MCE maximum temperature.

While the values $\Delta T_{ad}$ and $\Delta S_m$ in the moderate magnetic fields of 18 kOe field are large, in terms of practical application for magnetic refrigeration technology, these materials should not be considered promising, as the temperature width of half maximum is quite narrow (about 15 K) and the value of RCP is small respectively. In the 18 kOe field the RCP values are equal 61 and 70 J/kg for LCMO16 and LCMO18 respectively. Compared with classical magnetocaloric materials, these values are much smaller. But one can improve the magnetocaloric effectiveness of these materials, making sandwiches of LCMO16 and LCMO18. Magnetocaloric properties of these materials are almost identical, except for the temperature peaks, but these temperatures are also quite close to each other and are sub-room.

The sandwich of the LCMO16 and LCMO18 samples was produced for direct measurement of adiabatic temperature change $\Delta T_{ad}$. Samples with identical sizes were glued with each other, and the junction of chromel-constantan differential thermocouple was placed between the samples. Otherwise the experimental setup is similar to direct measurements of MCE by modulation method. Figure 5 shows experimental curve of $\Delta T_{ad}$ obtained for the sandwich. This figure also shows the $\Delta T_{ad}$ and $\Delta S_m$ curves for sandwiches calculated from the LCMO16 and LCMO18 curves. Masses of the samples were identical, and specific heat can be considered approximately equal, so for the calculation of the resulting curves, $\Delta T_{ad}$ for both samples were summed, and the values obtained were divided by two (and similarly for the $\Delta S_m$ calculation). It can be seen that the $\Delta T_{ad}$ values obtained this way coincide with high accuracy with the experimental values for the sandwich. Maximum values of $\Delta T_{ad}$ and $\Delta S_m$ for the sandwich are smaller than for pure LCMO16 and LCMO18, but the width of the effect is broadened. RCP for such material is 85 J/kg, which is higher than for LCMO16 or LCMO18 manganites. The similar increase of the RCP for a $La_{0.7}Ca_{0.3}MnO_3/La_{0.8}Sr_{0.2}MnO_3$ composite by means of magnetic measurements was observed by Pekala et al. [16].

Thus, the use of sandwich of materials with similar magnetocaloric properties increases RCP by about 20%. This method based on existing materials, is one for improving the efficiency of magnetic refrigerators, along with frequency increase of magnetization-demagnetization processes.

In conclusion it should be noted that direct measurement of MCE showed that the magnitude of the effect in $La_{0.7}Ca_{0.3}MnO_3$ reaches a considerable value, but the width of the temperature effect is quite narrow. The $^{16}O \rightarrow {}^{18}O$ isotope substitution shifts the temperature of MCE maximum to low temperatures without changing the MCE value. Production a sandwich of materials with close temperatures of MCE peaks can increase RCP compared with the initial materials.

**Acknowledgements**

This work was supported by t and the Branch of Physical Sciences of the Russian Academy of Sciences within the framework of the program "Strongly Correlated Electrons in Solids and Structures, RFBR (Grant Nos. 11-02-01124, 12-02-96506, 12-02-31171) and the State Contracts No. 16.552.11.7092 and 16.523.11.3008

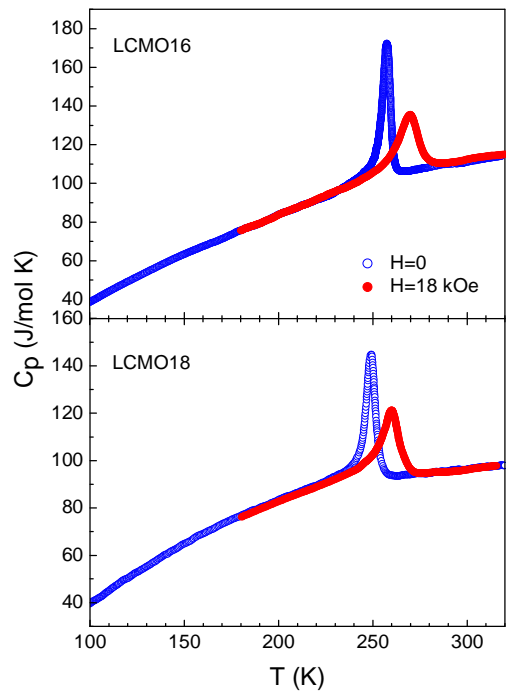

Figure 1. Specific heat of $La_{0.7}Ca_{0.3}Mn^{16}O_3$ (upper panel) and $La_{0.7}Ca_{0.3}Mn^{18}O_3$.

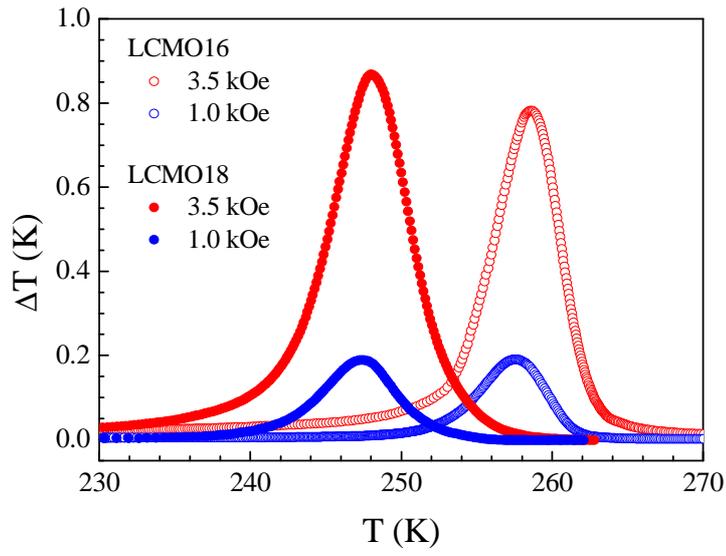

Figure 2 Magnetocaloric effect of $La_{0.7}Ca_{0.3}Mn^{16}O_3$ and $La_{0.7}Ca_{0.3}Mn^{18}O_3$ in low magnetic fields.

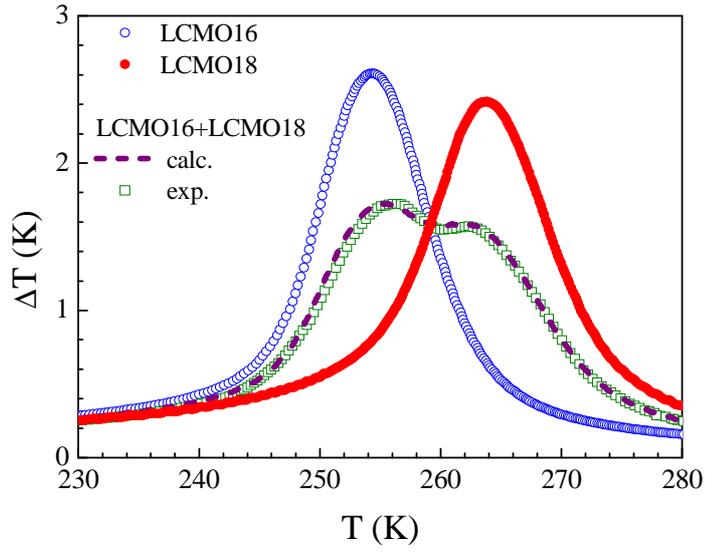

Figure 3. Magnetocaloric effect of $La_{0.7}Ca_{0.3}Mn^{16}O_3$, $La_{0.7}Ca_{0.3}Mn^{18}O_3$ and LCMO16+LCMO18 'sandwich' at $\Delta H$=18 kOe.

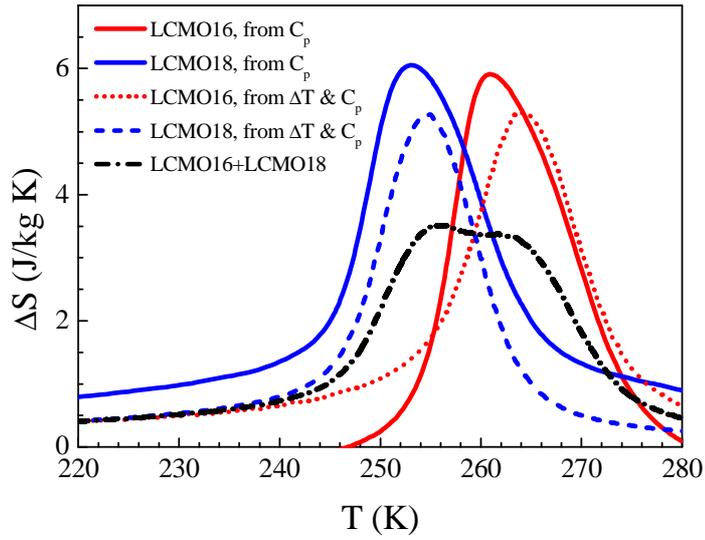

Figure 4. Magnetic entropy changes of $La_{0.7}Ca_{0.3}Mn^{16}O_3$ and $La_{0.7}Ca_{0.3}Mn^{18}O_3$ at $\Delta H$=18 kOe, calculated using specific heat data in and without magnetic field (solid lines) and specific heat and $\Delta T_{ad}$ data. Dash dot line – calculated $\Delta S_M$ for LCMO16+LCMO18 sandwich.